\documentclass[11pt,a4paper]{article}
\usepackage[hyperref]{acl2021}
\usepackage{times}
\usepackage{latexsym}

\usepackage{microtype}

\aclfinalcopy % Uncomment this line for the final submission
\usepackage{microtype}
\usepackage{times}
\usepackage{latexsym}
\usepackage{url}
\usepackage{multirow}
\usepackage{balance}
\usepackage{latexsym}
\usepackage{amsmath}
\usepackage{paralist}
\usepackage{makecell}
\usepackage{amssymb}
\usepackage{subcaption}
\usepackage{caption}
\usepackage{graphicx}

\usepackage{url}
\usepackage{multicol}
\usepackage{multirow}
\usepackage{booktabs}
\usepackage{color}
\usepackage{transparent}
\usepackage{array}
\usepackage{balance}
\usepackage{bm}
\usepackage{tabularx}

\definecolor{midnightgreen}{rgb}{0.0, 0.29, 0.33}
\definecolor{darkpink}{rgb}{0.91, 0.33, 0.5}
\definecolor{brown}{RGB}{119, 33, 6}

\newcommand{\rein}{\text{ReInfoSelect}}
\newcommand{\meta}{\text{MetaAdaptRank}}
\newcommand{\contrast}{\text{CTSyncSup}}
\newcommand{\QG}{\text{SyncSup}}

\newcommand{\redbf}[1]{\textcolor{red}{\bf {#1}}}
\newcommand{\bluebf}[1]{\textcolor{blue}{\bf {#1}}}

\title{Few-Shot Text Ranking with Meta Adapted Synthetic Weak Supervision}

\author{
\textbf{Si Sun}$^{1}$, \textbf{Yingzhuo Qian}$^{2}$, \textbf{Zhenghao Liu}$^2$, \textbf{Chenyan Xiong}$^3$,\\
\textbf{Kaitao Zhang}$^2$, \textbf{Jie Bao}$^{1}$, \textbf{Zhiyuan Liu}$^2$, \textbf{Paul Bennett}$^3$ \\
$^1$Department of Electronic Engineering, Tsinghua University, Beijing, China\\
$^2$Department of Computer Science and Technology, Tsinghua University, Beijing, China\\
Institute for Artificial Intelligence, Tsinghua University, Beijing, China\\
Beijing National Research Center for Information Science and Technology, China\\
$^3$Microsoft Research, Redmond, USA\\
\texttt{\{s-sun17, qyz17, liu-zh16, zkt18\}@mails.tsinghua.edu.cn}\\
\texttt{\{bao, liuzy\}@tsinghua.edu.cn}\\
\texttt{\{chenyan.xiong, Paul.N.Bennett\}@microsoft.com}
}

\date{}
\begin{document}
\maketitle

%% ********************************************
\begin{abstract}
The effectiveness of Neural Information Retrieval (Neu-IR) often depends on a large scale of in-domain relevance training signals, which are not always available in real-world ranking scenarios. To democratize the benefits of Neu-IR, this paper presents \meta{}, a domain adaptive learning method that generalizes Neu-IR models from label-rich source domains to few-shot target domains. Drawing on source-domain massive relevance supervision, \meta{} contrastively synthesizes a large number of weak supervision signals for target domains and meta-learns to reweight these synthetic ``weak'' data based on their benefits to the target-domain ranking accuracy of Neu-IR models. Experiments on three TREC benchmarks in the web, news, and biomedical domains show that \meta{} significantly improves the few-shot ranking accuracy of Neu-IR models. Further analyses indicate that \meta{} thrives from both its contrastive weak data synthesis and meta-reweighted data selection. The code and data of this paper can be obtained from~\url{https://github.com/thunlp/MetaAdaptRank}.
\end{abstract}

\section{Introduction}
Text retrieval aims to rank documents to either directly satisfy users' search needs or find textual information for later processing components, e.g., question answering~\cite{chen2017reading} and fact verification~\cite{liu2020fine}. Neural information retrieval (Neu-IR) models have recently shown advanced results in many ranking scenarios where massive relevance labels or clickthrough data are available~\cite{mitra2018introduction, craswell2020overview}.

The flip side is that the ``data-hungry'' nature of Neu-IR models yields mixed results in few-shot ranking scenarios that suffer from the shortage of labeled data and implicit user feedback~\cite{lin2019neural, yang2019critically}. On ranking benchmarks with only hundreds of labeled queries, there have been debates about whether Neu-IR, even with billions of pre-trained parameters~\cite{zhang2020rapidly}, really outperforms traditional IR techniques such as feature-based models and latent semantic indexing~\cite{yang2019critically, roberts2020trec}. In fact, many real-world ranking scenarios are few-shot, e.g., tail web queries that innately lack large supervision~\cite{downey2007heads}, applications with strong privacy constraints like personal and enterprise search~\cite{chirita2005using, hawking2004challenges}, and domains where labeling requires professional expertise such as biomedical and legal search~\cite{roberts2020trec, arora2018challenges}.

To broaden the benefits of Neu-IR to few-shot scenarios, we present an adaptive learning method \meta{} that meta-learns to adapt Neu-IR models to target domains with synthetic weak supervision. For synthesizing weak supervision, we take inspiration from the work~\cite{ma2021zero} that generates related queries for unlabeled documents in a zero-shot way, but we generate discriminative queries based on contrastive pairs of relevant (positive) and irrelevant (negative) documents. By introducing the negative contrast, \meta{} can subtly capture the difference between documents to synthesize more ranking-aware weak supervision signals. Given that synthetic weak supervision inevitably contains noises, \meta{} meta-learns to reweight these synthetic weak data and trains Neu-IR models to achieve the best accuracy on a small volume of target data. In this way, neural rankers can distinguish more useful synthetic weak supervision based on the similarity of the gradient directions of synthetic data and target data~\cite{ren2018learning} instead of manual heuristics or trial-and-error data selection~\cite{zhang2020selective}.

% To broaden the benefits of Neu-IR to few-shot scenarios, we present an adaptive learning method \meta{} that meta-learns to adapt Neu-IR models to target domains with synthetic weak supervision. For synthesizing weak supervision, we take inspiration from the work~\cite{ma2021zero} that generates related queries for unlabeled documents in a zero-shot way, but we generate discriminative queries based on contrastive pairs of relevant (positive) and irrelevant (negative) documents. By introducing the negative contrast, \meta{} can capture the subtle difference between documents to synthesize more ranking-aware weak relevance signals. Given that synthetic weak supervision inevitably contains noises, \meta{} meta-learns to reweight these synthetic data and trains Neu-IR models to achieve the best accuracy on a small volume of target data. In this way, neural rankers can distinguish more useful synthetic weak supervision based on the similarity of the gradient directions of synthetic data and target data~\cite{ren2018learning} instead of manual heuristics or trial-and-error data selection~\cite{zhang2020selective}.

We conduct experiments on three TREC benchmarks, ClueWeb09, Robust04, and TREC-COVID, which come from the web, news, and biomedical domains, respectively. \meta{} significantly improves the few-shot ranking accuracy of Neu-IR models across all benchmarks. We also empirically indicate that both contrastive weak data synthesis and meta-reweighted data selection contribute to \meta{}'s effectiveness. Compared to prior work~\cite{ma2021zero, zhang2020selective}, \meta{} not only synthesizes more informative queries and effective weak relevance signals but customizes more diverse and fine-grained weights on synthetic weak data to better adapt neural rankers to target few-shot domains.

%% **********************************
%%% Model Framework
\begin{figure*}[t]
  \centering
  \resizebox{\textwidth}{!}{
  \includegraphics[scale=0.505]{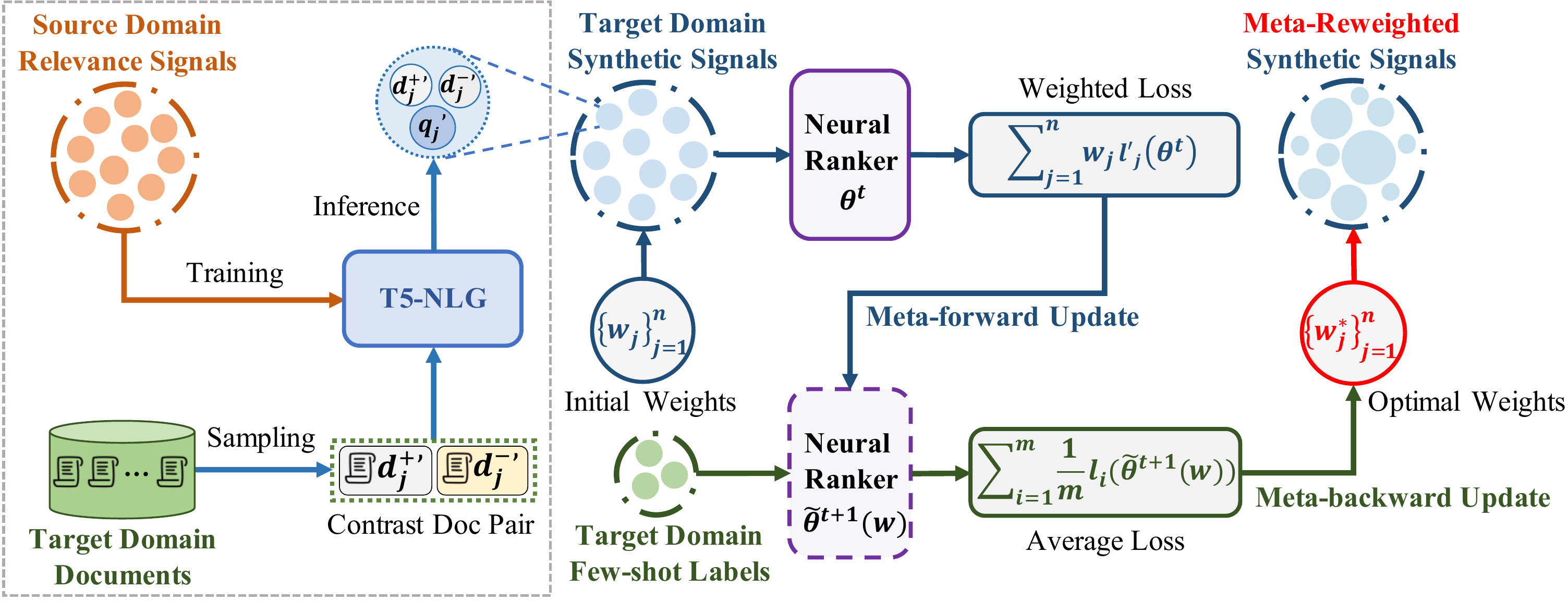}
  }
\caption{\label{fig:model_outline} The illustration of \meta{}, which first synthesizes massive weak supervision signals for target domains, and then meta-learns to reweight these synthetic data based on small target-domain relevance labels.}
\end{figure*}

\section{Related Work}
Recent Neu-IR methods have achieved promising results in modeling relevance matching patterns between queries and documents~\cite{jiafeng2016deep, hui2017pacrr, mitra2018introduction}. They have been extensively employed in ad-hoc text retrieval~\cite{xiong2017knrm, convknrm, nogueira2019passage, xiong2021approximate} and later natural language processing (NLP) tasks~\cite{lee2019latent, liu2020fine, qu2020open}.

The effectiveness of Neu-IR methods heavily relies on the end-to-end training with a large number of relevance supervision signals, e.g., relevance labels or user clicks. Nevertheless, such supervision signals are often insufficient in many ranking scenarios. The less availability of relevance supervision pushes some Neu-IR methods to freeze their embeddings to avoid overfitting~\cite{yates2020capreolus}. The powerful deep pre-trained language models, such as BERT~\cite{devlin2019bert}, also do not effectively alleviate the dependence of Neu-IR on a large scale of relevance training signals. Recent research even observes that BERT-based neural rankers might require more training data than shallow neural ranking models~\cite{hofsttter2020local, craswell2020overview}. Moreover, they may often be overly confident and more unstable in the learning process~\cite{qiao2019understanding}. 

% The effectiveness of Neu-IR methods heavily relies on the end-to-end training with a large amount of relevance supervision signals, e.g., relevance labels or user clicks. However, such relevance signals are often insufficient in many search scenarios. The less availability of supervision signals pushes some neural methods to freeze embeddings to avoid overfitting~\cite{yates2020capreolus}. The powerful deep pre-trained language models, such as BERT~\cite{devlin2019bert}, also do not effectively alleviate the dependence of Neu-IR on large-scale relevance supervision sources. Recent research even observes that BERT-based neural rankers might require more relevance training signals than shallow neural ranking models~\cite{hofsttter2020local, craswell2020overview}. Moreover, they may often be overly confident and more unstable in the learning process~\cite{qiao2019understanding}. 

A promising direction to alleviate the dependence of Neu-IR models on large-scale relevance supervision is to leverage weak supervision signals that are noisy but available at mass quantity~\cite{zheng2019investigating, dehghani2017neural, yu2020few}. Through IR history, various weak supervision sources have been used to approximate query-document relevance signals, e.g., pseudo relevance labels generated by unsupervised retrieval methods~\cite{dehghani2017neural, zheng2019investigating}, and title-document pairs~\cite{macavaney2019content}. Recently, \citet{zhang2020selective} treat paired anchor texts and linked pages as weak relevance signals and propose a reinforcement-based data selection method \rein{}, which learns to filter noisy anchor signals with trial-and-error policy gradients. Despite their convincing results, anchor signals are only available in web domains. Directly applying them to non-web domains may suffer from suboptimal outcomes due to domain gaps. To obtain weak supervision that adapts arbitrary domains, \citet{ma2021zero} present a synthetic query generation method, which can be trained with source-domain relevance signals and applied on target-domain documents to generate related queries.

More recently, a novel meta-learning technique has shown encouraging progress on solving data noises and label biases in computer vision~\cite{ren2018learning, shu2019meta, zheng2019meta} and some NLP tasks~\cite{zheng2019meta, wang2020balancing}. To the best of our knowledge, this novel technique has not been well utilized in information retrieval and synthetic supervision settings.

\section{Methodology}
This section first recaps the preliminary of Neu-IR and then introduces our proposed \meta{}. The framework of our method is shown in Figure~\ref{fig:model_outline}.

%% *************************************************
%% *************************************************
%% Neu-IR
\subsection{Preliminary of Neu-IR}
\label{sec:Neu-IR}
The ad-hoc retrieval task is to calculate a ranking score $f(q, d; \theta)$ for a query $q$ and a document $d$ from a document set. In Neu-IR, the ranking score $f(\cdot;\theta)$ is calculated by a neural model, e.g., BERT, with parameters $\theta$. The query $q$ and the document $d$ are encoded to the token-level representations $H$:
\begin{equation}
\small
H = \text{BERT}(\text{[CLS]} \circ q \circ \text{[SEP]} \circ d \circ \text{[SEP]}),
\end{equation}
where $\circ$ represents the concatenation operation. [CLS] and [SEP] are special tokens. The first token (``[CLS]'') representation $H_0$ is regarded as the representation of the $q$-$d$ pair. Then the ranking score $f(q, d; \theta)$ of the pair can be calculated as:
\begin{equation}
\small
f(q, d; \theta) = \text{tanh}(\text{Linear}(H_0)).
\end{equation}

The standard learning to rank loss $l_i(\theta)$~\cite{liu2009learning}, e.g., pairwise loss, can be used to optimize the neural model with relevance supervision signals $\{(q_i, d_i^+, d_i^-), 1\leq i \leq M\}$:
\begin{equation}
\small
l_i(\theta) = \text{relu}(1 - (f(q_i, d_i^+; \theta) - f(q_i, d_i^-; \theta))),
\end{equation}
where $d_i^+$ and $d_i^-$ denote the relevant (positive) and irrelevant (negative) documents of the query $q_i$. In few-shot ranking scenarios, the number of relevance supervision signals ($M$) is limited, making it difficult to train an accurate Neu-IR model. 

To mitigate the few-shot challenge in Neu-IR, \meta{} first transfers source-domain supervision signals to target-domain weak supervision signals (Sec~\ref{sec:ContrastQG}); then meta-learns to reweight the synthetic weak supervision (Sec~\ref{sec:Meta}) for selectively training Neu-IR models (Sec~\ref{sec:actual_train}).

%% *************************************************
%% *************************************************
%% ContrastQG
\subsection{Contrastive Synthetic Supervision}
\label{sec:ContrastQG}
\meta{} transfers the relevance supervision signals from source domains to few-shot target domains in a zero-shot way. In this way, a natural language generation (NLG) model is trained on source domain relevance signals (\textit{Source-domain NLG Training}) and is employed in target domains to synthesize weak supervision signals (\textit{Target-domain NLG Inference}). We will first recap the previous synthetic method~\cite{ma2021zero} and then introduce our contrastive synthetic approach.

\textbf{Preliminary of Synthetic Supervision.} Given a large volume of source-domain relevance pairs $(q, d^+)$, previous synthetic method~\cite{ma2021zero} trains a NLG model such as T5~\cite{raffel2020exploring} that learns to generate a query $q$ based on its relevant document $d^+$:
\begin{equation}
\small
q = \text{T5-NLG}(\text{[POS]} \circ d^+ \circ \text{[SEP]}), \label{eq:QG}
\end{equation}
where [POS] and [SEP] are special tokens. In inference, the trained query generator is directly used to generate new queries $q^*$ for target-domain documents $d^*$, where $d^*$ is regarded as the related (positive) document of $q^*$, while the unrelated (negative) document can be sampled from the target corpus.

Despite some promising results, the vanilla training strategy may cause the NLG model to prefer to generate broad and general queries that are likely related to a crowd of documents in the target corpus. As a consequence, the synthetic relevance supervision does not have enough ranking awareness to train robust Neu-IR models.

% Despite some promising results, the vanilla training strategy may cause the NLG model to prefer to generate broad and general queries that are likely related to a crowd of target domain documents. This might result in the lack of ranking awareness of the synthetic weak supervision signals, making it hard to train enough robust Neu-IR models.

% \textbf{Source-domain NLG Training.} To synthesize ranking-aware weak supervision, \meta{} trains the NLG model such that captures the subtle difference between the contrastive document pair ($d^+$, $d^-$) to generate a discriminative query $q$:

\textbf{Source-domain NLG Training.} To synthesize ranking-aware weak supervision, \meta{} trains the NLG model to capture the difference between the contrastive document pair ($d^+$, $d^-$) and generate a discriminative query $q$:
\begin{equation}
\small
q = \text{T5-NLG}(\text{[POS]} \circ d^+ \circ \text{[NEG]} \circ d^- \circ \text{[SEP]}), \label{eq:ContrastQG}
\end{equation}
where [NEG] is another special token. The training instances $(q, d^+, d^-)$ can be obtained from source domains in which $d^+$ and $d^-$ are annotated as the relevant and irrelevant documents for the query $q$.

\textbf{Target-domain NLG Inference.} During inference, we first pick out a mass of confusable document pairs from target domains and then feed them into our trained contrastive query generator (Eq.~\ref{eq:ContrastQG}) to synthesize more valuable weak supervision data.

To get confusable document pairs, we first generate a seed query $q^*$ for each target-domain document $d^*$ using the trained query generator (Eq.~\ref{eq:QG}). Then the seed query is used to retrieve a subset of documents with BM25, where other retrieval methods can also be utilized. The confusable document pairs (${d^+}'$, ${d^-}'$) are pairwise sampled from the retrieved subset without considering their rankings. Given the confusable document pair, we leverage our trained contrastive query generator to generate a new query ${q}'$:
\begin{equation}
\small
{q}' = \text{T5-NLG}(\text{[POS]} \circ {d^+}' \circ \text{[NEG]} \circ {d^-}' \circ \text{[SEP]}),
\end{equation}
where ${d^+}'$ and ${d^-}'$ are regarded as the related (positive) and unrelated (negative) documents of ${q}'$. In this way, we can synthesize massive target-domain weak supervision $\{({q_j}', {d_j^+}', {d_j^-}'), 1\leq j \leq N\}$.

%% *************************************************
%% *************************************************
%% Meta Learning to Reweight
\subsection{Meta Learning to Reweight}
\label{sec:Meta}

The synthetic weak data inevitably contain noises. To distinguish more useful training data for neural rankers, \meta{} meta-learns to reweight these synthetic data, following~\citet{ren2018learning}. 

% The synthetic weak data inevitably contain noises. To distinguish more useful training data for neural rankers, \meta{} meta-learns to reweight these synthetic data, following~\citet{ren2018learning}. 

\textbf{Meta Learning Objective.} Given a large volume of synthetic data $\{({q_j}', {d_j^+}', {d_j^-}'), 1\leq j \leq N\}$ and a handful of target data $\{(q_i, d_i^+, d_i^-), 1\leq i \leq M\}$ ($M \ll N$), our meta-learning objective is to find the optimal weights $w^{*}$ on synthetic data to better train neural rankers. The learning of $w^{*}$ involves \textit{two nested loops of optimization}: initial-weighted synthetic data is used to \textit{pseudo-optimize} the neural ranker; the weights is then \textit{optimized} by minimizing the neural ranking loss on target data. 

% Specifically
To be specific, the first loop (\textit{Meta-forward Update}) incorporates the initial weights $w$ into the learning parameters $\widetilde{\theta}(w)$ instead of truly optimizing the neural ranker:
\begin{equation}
    \small
    \widetilde{\theta}(w) = \arg\min_{\theta}  \sum_{j=1}^N w_j l'_j (\theta), \label{eq:weak}
\end{equation}
where $l'_j (\theta)$ is the ranking loss on a synthetic instance $({q_j}', {d_j^+}', {d_j^-}')$. In the second loop (\textit{Meta-backward Update}), the optimal weights $w^*$ can be obtained by minimizing the target ranking loss:
\begin{equation}
    \small
    w^{*} = \arg\min_{w} \sum_{i=1}^M l_i(\widetilde{\theta}(w)), \label{eq:meta}
\end{equation}
where $l_i (\theta)$ is the ranking loss on a target instance $(q_i, d_i^+, d_i^-)$. The calculation of each loop can be very expensive. In practice, we only perform one-step optimization in the two loops with mini-batch data, consistent with prior work~\cite{ren2018learning}.

\textbf{Meta-forward Update.} Taking the $t$-th training step as an example, we first assign a set of initial weights $w = \{w_j\}_{j=1}^{n}$ to the synthetic training data batch and then pseudo-update the neural ranker's parameters to $\widetilde{\theta}^{t+1}(w)$:
\begin{equation}
    \small
    \widetilde{\theta}^{t+1}(w)  = \theta^{t} - \alpha \frac{\partial  }{\partial  (\theta^{t})} \sum_{j=1}^n w_j l'_j(\theta^t), \label{eq:meta-forward}
\end{equation}
where $\alpha$ is the learning rate. The description here uses vanilla SGD and other optimizers can be used.

\textbf{Meta-backward Update.} We leverage the neural ranker $\widetilde{\theta}^{t+1}(w)$ to calculate the ranking loss on the target data batch and obtain the optimal weights $w^{*} = \{w_j^{*}\}_{j=1}^{n}$ through a single optimization step:
\begin{equation}
    \small
    w^{*}_{j} =  w_j - \eta \frac{\partial  }{\partial  (w_j)}  \sum_{i=1}^{m} \frac{1}{m} l_i (\widetilde{\theta}^{t+1}(w)), \label{eq.meta-gradient}
\end{equation}
where $\eta$ is the learning rate for optimizing weights. The weights are further normalized for stable training. More details are shown in Appendices~\ref{sec:appendix:meta}.

\subsection{Training with Meta-Weights}
\label{sec:actual_train}

After obtaining the optimal weights $w^{*}$, the optimization of the neural ranker is a standard back-propagation on the weighted loss of synthetic data:
\begin{equation}
\small
    \theta^{t+1} = \arg\min_{\theta^t} \sum_{j=1}^n w_j^{*} l'_j(\theta^t).
\end{equation}
In each training step, \meta{} first learns to reweight the synthetic batch based on their meta-impact on the target batch and then updates the neural ranker with the weighted synthetic batch. In this way, the few-shot target data can serve more as a ``regularizer'' to help the neural ranker to generalize with synthetic data, instead of as direct supervision which requires more labels~\cite{ren2018learning}.

\section{Experimental Methodology}
~\label{sec:method}

This section describes our experimental settings and implementation details.

% ******************************************
% ******************************************
%%% Datasets
\textbf{Datasets.} As shown in~Table~\ref{tab:dataset}, three standard TREC datasets with different domains are used in our experiments: ClueWeb09-B~\cite{callan2009clueweb09}, Robust04~\cite{kwok2004trec}, and TREC-COVID~\cite{roberts2020trec}. They are all few-shot ad-hoc retrieval datasets where the number of labeled queries is limited. We leverage the ``Complete" version of TREC-COVID whose retrieval document set is the July 16, 2020 release of CORD-19~\cite{wang2020cord}, a growing collection of scientific papers on COVID-19 and related research.

% ******************************************
% ******************************************
\textbf{Evaluation Settings.} We evaluate supervised IR methods through re-ranking the top 100 documents from the first-stage retrieval with five-fold cross-validation, consistent with prior work~\cite{xiong2017duet, dai2019deeper, zhang2020selective}. The first-stage retrieval for ClueWeb09-B and Robust04 is the sequential dependence model (SDM)~\cite{metzler2005markov} released by \citet{dai2019deeper}, and the first-stage retrieval for TREC-COVID is BM25~\cite{robertson2009probabilistic} well-tuned by Anserini~\cite{yang2017anserini}.

% \footnote{\url{https://github.com/castorini/anserini}}. 

% ******************************************
% ******************************************
\textbf{Metrics.} NDCG@20 is used as the primary metric for all datasets. We also report ERR@20 for ClueWeb09-B and Robust04, which is the same with prior work~\cite{zhang2020selective}, and report P@20 for TREC-COVID. Statistic significance is examined by permutation test with $p<0.05$.

% ******************************************
% ******************************************
%% Baselines
\textbf{Baselines.} Two groups of baselines are compared in our experiments, including \textit{Traditional IR Baselines} and \textit{Neural IR Baselines}.

% ******************************************
%% Traditional IR Baselines
\textit{Traditional IR Baselines.} Following previous research~\cite{dai2019deeper, zhang2020selective}, we compare four traditional IR methods in our experiments. They are two unsupervised methods, \texttt{BM25}~\cite{robertson2009probabilistic} and \texttt{SDM}~\cite{metzler2005markov}, and two learning-to-rank (LTR) methods using bag-of-word features, \texttt{RankSVM}~\cite{joachims2002optimizing} and \texttt{Coor-Ascent} (Coordinate Ascent)~\cite{metzler2007linear}.

% ******************************************
%% Neu-IR Baselines
\textit{Neural IR Baselines.} We also compare seven Neu-IR baselines that utilize different methodologies to train neural rankers. In our experiments, all Neu-IR methods adopt the widely-used BERT ranker~\cite{nogueira2019passage}, BERT-FirstP, which only uses the first paragraph of documents.

The vanilla neural baseline only leverages the existing small-scale relevance labels of target datasets to train BERT rankers, which is named \texttt{Few-shot Supervision}. We also compare BERT rankers trained with two large-scale supervision sources: \texttt{Bing User Click} and \texttt{MS MARCO}. \citet{dai2019deeper} train BERT rankers with 5 million user click logs in Bing. We borrow their reported results because commercial logs are not publicly available. MS MARCO is a human supervision source~\cite{nguyen2016ms}, which provides over one million Bing queries with relevance labels. 

Four weak supervision methods are also compared. One baseline is \texttt{Title Fitler}, which treats filtered title-document pairs as weak supervision signals~\cite{macavaney2019content} for training BERT rankers~\cite{zhang2020selective}. Another two baselines are \texttt{Anchor} and \texttt{\rein{}}. \texttt{Anchor} leverages 100k pairs of anchor texts and web pages to train BERT rankers~\cite{zhang2020selective}. \texttt{\rein{}} first employs reinforcement learning to select these anchor signals~\cite{zhang2020selective} and then trains BERT rankers. The last baseline \texttt{\QG{}} trains BERT rankers with synthetic weak supervision data, which are synthesized based on the previous work~\cite{ma2021zero}.

%% ********************************
\begin{table}[t]
\centering
\small
% \scalebox{0.8}{
\resizebox{\columnwidth}{!}{
\begin{tabular}{l c c c}
  \toprule
  \textbf{Dataset} & \textbf{Domain} & \textbf{Corpus Size} & \textbf{Labeled Queries} \\
  \hline
  ClueWeb09-B & Web Pages & 50m & 200 \\
  Robust04 & News Articles & 528k & 250 \\
  TREC-COVID & BioMed Papers & 191k & 50 \\
  \bottomrule
  \end{tabular}
  }
  \caption{\label{tab:dataset}Statistics of three TREC datasets used in our experiments. They are few-shot ranking datasets containing only tens to hundreds of labeled queries.}
\end{table}

%  
% 
% 
% \begin{table}[t]
% \centering
% \small
% \scalebox{0.8}{
% \begin{tabular}{c c c c}
%   \toprule
%   \textbf{Dataset} & \textbf{Domain} & \textbf{Documents} & \textbf{Queries (labeled)} \\
%   \hline
%   ClueWeb09-B & Web Pages & 50m & 200 \\
%   Robust04 & News Articles & 528k & 249 \\
%   TREC-COVID & BioMed Papers & 191k & 50 \\
%   \bottomrule
%   \end{tabular}
%   }
%   \caption{\label{tab:dataset}Statistics of few-shot testing benchmarks where only dozens to hundreds of labeled queries.}
% \end{table}

% \begin{table}[t]
% \centering
% \small
% \scalebox{0.8}{
% \begin{tabular}{c c c c c}
%   \toprule
%   \textbf{Collection} & \textbf{Domain} & \textbf{Queries} &  \textbf{Documents} & \textbf{Labels} \\
%   \hline
%   ClueWeb09-B & Web Pages & 200 & 50m & 47121 \\
%   Robust04 & News Articles & 249 & 528k & 311410 \\
%   TREC-COVID & BioMed Papers & 50 & 191k & 69317 \\
%   \bottomrule
%   \end{tabular}
%   }
%   \caption{\label{tab:dataset}Statistics of few-shot testing benchmarks where only dozens to hundreds of labeled queries.}
% \end{table}
%% ********************************

%% ********************************
\begin{table*}[t]
\centering
\small
\resizebox{\textwidth}{!}{
  \begin{tabular}{l|l l|l l|l l}
  \toprule 
  \multirow{2}{*}{\textbf{Methods}} & \multicolumn{2}{c|}{\textbf{ClueWeb09-B (Web)}} & \multicolumn{2}{c|}{\textbf{Robust04 (News)}} & \multicolumn{2}{c}{\textbf{TREC-COVID (BioMed)}} \\
   & \textbf{NDCG@20} & \textbf{ERR@20} & \textbf{NDCG@20} & \textbf{ERR@20} & \textbf{NDCG@20} & \textbf{P@20} \\
   \hline
   BM25~\cite{yang2017anserini} & 0.2773 & 0.1426 & 0.4129 & 0.1117 & 0.6979 & 0.7670 \\
   SDM~\cite{dai2019deeper} & 0.2774 & 0.1380 & 0.4269 & 0.1172 & 0.7030 & 0.7770 \\
   RankSVM~\cite{dai2019deeper} & 0.289 & n.a. & 0.420 & n.a. & n.a.  & n.a.  \\
   RankSVM (OpenMatch) & 0.2825 & 0.1476 & 0.4309 & 0.1173 & 0.6995  & 0.7570  \\
   Coor-Ascent~\cite{dai2019deeper} & 0.295 & n.a. & 0.427 & n.a. & n.a. & n.a. \\
   Coor-Ascent (OpenMatch) & 0.2969$\text{}^{\dagger}$ & 0.1581$\text{}^{\dagger}$ & 0.4340$\text{}^{\dagger}$ & 0.1171 & 0.7041 & 0.7770 \\
   \hline
   Few-shot Supervision~\cite{zhang2020selective} & 0.2999 & 0.1631 & 0.4258 & 0.1163 & n.a. & n.a \\ 
   Few-shot Supervision (Ours) & 0.3033$\text{}^{\dagger}$ & 0.1519 & 0.4572$\text{}^{\dagger \ddagger}$ & 0.1234 & 0.7713$\text{}^{\dagger \ddagger}$ & 0.8400$\text{}^{\dagger \ddagger}$ \\
   
   Bing User Click~\cite{dai2019deeper} & 0.333 & n.a. & n.a. & n.a. & n.a. & n.a. \\
   
   MS MARCO~\cite{nguyen2016ms} & 0.3205$\text{}^{\dagger \ddagger \flat \S}$ & 0.1690$\text{}^{\dagger \flat}$ & 0.4674$\text{}^{\dagger \ddagger}$ & 0.1304$\text{}^{\dagger \ddagger \flat}$ & 0.8054$\text{}^{\dagger \ddagger \flat}$ & 0.8610$\text{}^{\dagger \ddagger \flat}$ \\
   
   Title Filter~\cite{macavaney2019content} & 0.3021 & 0.1513 & 0.4379 & 0.1202 & n.a. & n.a. \\
   
   Anchor~\cite{zhang2020selective} & 0.3072$\text{}^{\dagger}$ & 0.1609$\text{}^{\dagger}$ & 0.4449$\text{}^{\dagger \ddagger}$ & 0.1223$\text{}^{\dagger \ddagger}$ & 0.7677$\text{}^{\dagger \ddagger}$ & 0.8260$\text{}^{\dagger \ddagger}$ \\
   
   ReInfoSelect~\cite{zhang2020selective} & 0.3261$\text{}^{\dagger \ddagger \flat \S}$ & 0.1669$\text{}^{\dagger \flat}$ & 0.4703$\text{}^{\dagger \ddagger \flat}$ & 0.1313$\text{}^{\dagger \ddagger \flat}$ & 0.7833$\text{}^{\dagger \ddagger}$ & 0.8420$\text{}^{\dagger \ddagger}$ \\
   
   \QG~\cite{ma2021zero} & 0.3036$\text{}^{\dagger}$ & 0.1602$\text{}^{\dagger}$ & 0.4685$\text{}^{\dagger \ddagger}$ & 0.1311$\text{}^{\dagger \ddagger \flat}$ & 0.7867$\text{}^{\dagger \ddagger}$ & 0.8470$\text{}^{\dagger \ddagger}$ \\
   \hline
   
   \contrast{} & 0.3123$\text{}^{\dagger}$ & 0.1764$\text{}^{\dagger \flat \S}$ & 0.4769$\text{}^{\dagger \ddagger \flat}$ & 0.1293$\text{}^{\dagger \ddagger \flat}$ & 0.8006$\text{}^{\dagger \ddagger \flat}$ & 0.8610$\text{}^{\dagger \ddagger}$ \\
   
   \meta{} & \textbf{0.3416}$\text{}^{\dagger \ddagger \flat \natural \S}$ & \textbf{0.1893}$\text{}^{\dagger \ddagger \flat \natural \sharp \S}$ & \textbf{0.4916}$\text{}^{\dagger \ddagger \flat \natural \sharp \S}$ & \textbf{0.1362}$\text{}^{\dagger \ddagger \flat \natural \S}$ & \textbf{0.8378}$\text{}^{\dagger \ddagger \flat \natural \sharp \S}$ & \textbf{0.8790}$\text{}^{\dagger \ddagger \flat \sharp \S}$ \\
    \bottomrule
  \end{tabular}
  }
  \caption{\label{tab:overall}Ranking accuracy of \meta{} and baselines. $\dagger, \ddagger, \flat, \natural, \sharp, \S$ indicate statistically significant improvements over SDM$\text{}^{\dagger}$, Coor-Ascent$\text{}^{\ddagger}$, Few-shot Supervision$\text{}^{\flat}$, MS MARCO$\text{}^{\natural}$, ReInfoSelect$\text{}^{\sharp}$ and \QG$\text{}^{\S}$.}
\end{table*}

\begin{table*}[ht]
\centering
\small
\scalebox{0.9}{
  \begin{tabular}{l l|l l|l l|l l}
  \toprule
  \multicolumn{2}{l|}{\multirow{2}{*}{\textbf{Supervision Sources}}} & \multicolumn{2}{c|}{\textbf{ClueWeb09-B (Web)}} & \multicolumn{2}{c|}{\textbf{Robust04 (News)}} & \multicolumn{2}{c}{\textbf{TREC-COVID (BioMed)}} \\
  & & \textbf{NDCG@20} & \textbf{ERR@20} & \textbf{NDCG@20} & \textbf{ERR@20} & \textbf{NDCG@20} & \textbf{P@20} \\
  \hline
  \textit{(a)} & MS MARCO~\cite{nguyen2016ms} & 0.3205$\text{}^{\flat}$ & 0.1690 & 0.4674$\text{}^{\ddagger}$ & 0.1304$\text{}^{\ddagger}$ & 0.8054$\text{}^{\ddagger}$ & 0.8610$\text{}^{\ddagger}$ \\
  
  \textit{(b)} & Anchor~\cite{zhang2020selective} & 0.3072 & 0.1609 & 0.4449 & 0.1223 & 0.7677 & 0.8260 \\
  
  \textit{(c)} & SyncSup~\cite{ma2021zero} & 0.3036 & 0.1602 & 0.4685$\text{}^{\ddagger}$ & \textbf{0.1311}$\text{}^{\ddagger}$ & 0.7867 & 0.8470 \\
  
  \hline
  \textit{(d)} & \contrast{} & 0.3123 & \textbf{0.1764}$\text{}^{\flat}$ & \textbf{0.4769}$\text{}^{\ddagger}$ & 0.1293$\text{}^{\ddagger}$ & 0.8006$\text{}^{\ddagger}$ & 0.8610$\text{}^{\ddagger}$ \\
  
  \textit{(e)} & MARCO + \contrast{} & \textbf{0.3214}$\text{}^{\flat}$ & 0.1739$\text{}^{\ddagger \flat}$ & 0.4727$\text{}^{\ddagger}$ & 0.1297$\text{}^{\ddagger}$ & \textbf{0.8182}$\text{}^{\ddagger \flat}$ & \textbf{0.8720}$\text{}^{\ddagger \flat}$ \\
  
  \bottomrule
  \end{tabular}
  }
  \caption{\label{tab:weak_data}Ranking accuracy with different supervision sources. MARCO + \contrast{} denotes the hybrid source of MS MARCO and \contrast{}. $\dagger, \ddagger, \flat$ indicate statistically significant improvements over \textit{(a)}$\text{}^{\dagger}$, \textit{(b)}$\text{}^{\ddagger}$ and \textit{(c)}$\text{}^{\flat}$.}
\end{table*}

\begin{table*}[h]
\centering
\small
\scalebox{0.9}{
  \begin{tabular}{l c c c c c c c c}
  \toprule 
  \textbf{Synthetic Methods} & \textbf{BLEU-1} & \textbf{BLEU-2} &  \textbf{ROUGE-1} & \textbf{ROUGE-2} & \textbf{ROUGE-L} & \textbf{NIST@1} & \textbf{NIST@2} & \textbf{METEOR} \\ 
  \hline
  \QG{}~\cite{ma2021zero} & 0.5672 & 0.4527  & 0.5928 & 0.3764 & 0.5745 & 5.8070 & 7.3315 & 0.3089 \\
  Reverse-\contrast{} & 0.3185 & 0.1807 & 0.3528 & 0.1088 & 0.3395 & 3.0076 &  3.3665 & 0.1610 \\
  \contrast{} & \textbf{0.5909} & \textbf{0.4627} & \textbf{0.6238} & \textbf{0.3844}  & \textbf{0.5955} & \textbf{6.1282} & \textbf{7.6314} & \textbf{0.3191} \\
  \bottomrule 
\end{tabular}
  }
  \caption{\label{tab:bleu_score}Evaluation results of the queries generated by different synthetic methods. In Reverse-\contrast{}, we swap the encoding order of contrastive document pairs, using original negative documents as positive documents.}
\end{table*}

\textbf{Implementation Details.} This part introduces the implement details of our method and baselines.

\textit{BERT Ranker.} For our methods and all Neu-IR baselines, we use the base version of BERT~\cite{devlin2019bert} on ClueWeb09-B and Robust04, and PubMedBERT (Base)~\cite{gu2020domain} on TREC-COVID. We leverage the \text{OpenMatch}~\cite{liu2021openmatch} implementation and obtain the pre-trained weights from Hugging Face~\cite{wolf2020transformers}.

% 
% 
% \footnote{\url{https://github.com/thunlp/OpenMatch}}

For all Neu-IR methods, we first use additional supervision sources such as weak supervision signals to train BERT rankers (except for \texttt{Few-shot Supervision}); then fine-tune the BERT rankers with the training folds of target datasets in the cross-validation. Following prior work~\cite{dai2019deeper, zhang2020selective}, the ranking features ([CLS] embeddings) of BERT are combined with the first-stage retrieval scores using Coor-Ascent for ClueWeb09-B and Robust04. We set the max input length to 512 and use Adam optimizer with a learning rate of 2e-5 and a batch size of 8.

\textit{Contrastive Supervision Synthesis.} We use the small version of T5 (60 million parameters) as the NLG models in \meta{}, and leverage MS MARCO as the training data for T5-NLG models. We set the maximum input length to 512 and use Adam to optimize the T5-NLG models with a learning rate of 2e-5 and a batch size of 4. In inference, the T5-NLG models are applied on target datasets with greedy search. Additionally, we consider \texttt{\contrast{}} as our ablation baseline, which directly trains BERT rankers on contrastive synthetic supervision data without meta-reweighting.

\textit{Meta Learning to Reweight.} The training folds of the target dataset are used as target data to guide the meta-reweighting to synthetic data. We set the batch size of synthetic data ($n$) and target data ($m$) to 8. The second-order gradient of the target ranking loss with regard to the initial weight (Eq.~\ref{eq.meta-gradient}) is implemented using the automatic differentiation in PyTorch~\cite{paszke2017automatic}.

\section{Evaluation Results}

In this section, we present the evaluation results of \meta{} and conduct a series of analyses and case studies to study its effectiveness.

\subsection{Overall Accuracy}
\label{sec:overall}

The ranking results of \meta{} and baselines are presented in Table~\ref{tab:overall}. 

On all benchmarks and metrics, \meta{} outperforms all baselines stably. Compared to the best feature-based LeToR method, Coor-Ascent, \meta{} outperforms it by more than 15\%. \meta{} even outperforms the strong Neu-IR baselines supervised with Bing User Click and MS MARCO, which demonstrates its effectiveness.

Specifically, \contrast{} directly improves the few-shot ranking accuracy of BERT rankers by 3\% on all benchmarks. In comparison to other weak supervision sources, filtered title-document relations, Anchor and \QG{}, \contrast{} shows more stable effectiveness across different benchmarks, revealing its domain-adaption advantages. Moreover, meta-reweighting \contrast{} brings further improvement and helps \meta{} outperform the latest selective Neu-IR method \rein{}. 

Next, we go ahead to analyze \meta{}'s contrastive synthesis and meta-reweighting.

%% *********************************
\begin{table*}[t]
\centering
\small
\scalebox{0.88}{
  \begin{tabular}{l l|l l|l l|l l}
%   \hline
  \toprule
  \multicolumn{2}{l|}{\multirow{2}{*}{\textbf{Methods (Supervision Sources)}}} & \multicolumn{2}{c|}{\textbf{ClueWeb09-B (Web)}} & \multicolumn{2}{c|}{\textbf{Robust04 (News)}} & \multicolumn{2}{c}{\textbf{TREC-COVID (BioMed)}} \\
 & & \textbf{NDCG@20} & \textbf{ERR@20} & \textbf{NDCG@20} & \textbf{ERR@20} & \textbf{NDCG@20} & \textbf{P@20} \\
  \hline
  \textit{(a)} & ReInfoSelect (MS MARCO) & 0.3294 & 0.1760 & 0.4756 & 0.1291 & 0.8229$\text{}^{\ddagger}$ & 0.8780$\text{}^{\ddagger}$ \\
  
  \textit{(b)} & ReInfoSelect (Anchor) & 0.3261 & 0.1669 & 0.4703 & 0.1313 & 0.7891 & 0.8430 \\
  
  \textit{(c)} & ReInfoSelect (\contrast{}) & 0.3243 & 0.1742 & 0.4816$\text{}^{\ddagger}$ & 0.1334 & 0.8230$\text{}^{\ddagger}$ & 0.8800$\text{}^{\ddagger}$ \\
  
  \hline
  \textit{(d)} & \meta{} (MS MARCO) & 0.3453$\text{}^{\dagger \ddagger \flat}$ & \textbf{0.2018}$\text{}^{\dagger \ddagger \flat \sharp}$ & 0.4853$\text{}^{\ddagger}$ & 0.1331 & 0.8354$\text{}^{\ddagger \sharp}$ & 0.8730$\text{}^{\ddagger}$ \\
  
  \textit{(e)} & \meta{} (Anchor) & 0.3374 & 0.1730 & 0.4797 & 0.1314 & 0.8045 & 0.8650 \\

  \textit{(f)} & \meta{} (\contrast{}) & 0.3416$\text{}^{\flat}$ & 0.1893$\text{}^{\ddagger \sharp}$ & 0.4916$\text{}^{\dagger \ddagger \sharp}$ & 0.1362$\text{}^{\dagger \sharp}$ & 0.8378$\text{}^{\ddagger \sharp}$ & 0.8790$\text{}^{\ddagger}$ \\
  \hline
  \textit{(g)} & \meta{} (MARCO + \contrast{}) & \textbf{0.3498}$\text{}^{\dagger \ddagger \sharp}$ & 0.1926$\text{}^{\ddagger \flat \sharp}$ & \textbf{0.4989}$\text{}^{\dagger \ddagger \flat \natural \sharp}$ & \textbf{0.1366}$\text{}^{\dagger \natural}$ & \textbf{0.8488}$\text{}^{\dagger \ddagger \flat \natural \sharp}$ & \textbf{0.8910}$\text{}^{\ddagger \natural \sharp}$ \\
%   \hline
  \bottomrule
  \end{tabular}
  } 
  \caption{\label{tab:adapt_ndcg}Ranking accuracy of \rein{} and \meta{} using different supervision sources. Superscripts $\dagger, \ddagger, \flat, \natural, \sharp, \S$ indicate statistically significant improvements over \textit{(a)}$\text{}^{\dagger}$, \textit{(b)}$\text{}^{\ddagger}$, \textit{(c)}$\text{}^{\flat}$, \textit{(d)}$\text{}^{\natural}$, \textit{(e)}$\text{}^{\sharp}$ and \textit{(f)}$\text{}^{\S}$.}
\end{table*}

\subsection{Effectiveness of Contrastive Synthesis}
\label{sec:contrast_analysis}

We analyze contrastive synthesis's effectiveness by its effect on ranking results and synthetic quality.

Table~\ref{tab:weak_data} presents the ranking accuracy based on our \contrast{} and four other supervision sources. \contrast{} outperforms Anchor and \QG{} stably across all datasets. On Robust04, \contrast{} even shows better performance than MS MARCO human labels. Besides, combining the sources of MS MARCO and \contrast{} can further improve the ranking accuracy on ClueWeb09-B and TREC-COVID, revealing that \contrast{} provides useful supervision signals applicable to various domains.

We further evaluate the quality of the queries generated in \QG{} and our \contrast{}, which are both synthetic methods for generating queries based on target documents. Following previous research~\cite{ma2021zero, yu2020few, celikyilmaz2020evaluation}, eight auto evaluation metrics are used in our evaluation. As shown in Table~\ref{tab:bleu_score}, \contrast{} outperforms \QG{} on all metrics. The results demonstrate that the contrastive pair of positive and negative documents does help the NLG model better approximate the golden queries. In addition, reversing the encoding order of the contrastive document pair causes a dramatic decrease in all evaluation scores of the generated queries. This further shows that our contrastive query generator can extract more specific and representative information from the positive documents, thereby generating more discriminative queries.

% reversing the encoding order of the contrastive document pair causes all evaluation scores of the generated queries to decrease dramatically. This further demonstrates that our contrastive query generator can extract more characteristic rather than general information from the positive documents to generate more discriminative queries.

%% *********************************
% \begin{table*}[ht]
\begin{table*}[t]
\centering
\small
\scalebox{0.9}{
\begin{tabular}{l l l l l l l l}
\toprule
\multicolumn{2}{l}{\multirow{2}{*}{\textbf{TREC-COVID R5 Methods}}} & \multicolumn{2}{c}{\textbf{All Queries}} & \multicolumn{2}{c}{\textbf{Old Queries}} & \multicolumn{2}{c}{\textbf{New Queries}}  \\
&  & \textbf{NDCG@20} & \textbf{P@20} & \textbf{NDCG@20} & \textbf{P@20} & \textbf{NDCG@20} & \textbf{P@20} \\ 
\hline
\textit{(a)} & r5.fusion1 (Anserini BM25)  & 0.5313 & 0.5840 & 0.5202 & 0.5722 & 0.6320 & 0.6900 \\

\textit{(b)} & r5.fusion2 (Anserini BM25)  & 0.6007$\text{}^{\dagger}$ & 0.6440$\text{}^{\dagger}$ & 0.5937$\text{}^{\dagger}$ & 0.6344$\text{}^{\dagger}$ & 0.6641 & 0.7300 \\

\textit{(c)} & covidex.r5.2s (RRF)    & 0.7457$\text{}^{\dagger \ddagger}$  &  0.7610$\text{}^{\dagger \ddagger}$ & 0.7303$\text{}^{\dagger \ddagger}$ & 0.7456$\text{}^{\dagger \ddagger}$ & 0.8837$\text{}^{\dagger}$ & 0.9000  \\

\textit{(d)} & \meta{} (rerank \textit{(a)}) & 0.7536$\text{}^{\dagger \ddagger}$ & 0.7820$\text{}^{\dagger \ddagger}$ & 0.7405$\text{}^{\dagger \ddagger}$ & 0.7656$\text{}^{\dagger \ddagger}$ & 0.8712$\text{}^{\dagger \ddagger}$ & 0.9300$\text{}^{\dagger \ddagger}$  \\

\textit{(e)} & covidex.r5.d2q.2s (RRF) & 0.7539$\text{}^{\dagger \ddagger}$ & 0.7700$\text{}^{\dagger \ddagger}$ & 0.7385$\text{}^{\dagger \ddagger}$ & 0.7544$\text{}^{\dagger \ddagger}$ &  0.8929$\text{}^{\dagger}$ & 0.9100 \\

\textit{(f)} & \meta{} (rerank \textit{(b)})  & 0.7904$\text{}^{\dagger \ddagger \flat \natural}$ & 0.8270$\text{}^{\dagger \ddagger \flat \natural \sharp}$ & 0.7790$\text{}^{\dagger \ddagger \flat \natural}$ & 0.8144$\text{}^{\dagger \ddagger \flat \natural \sharp}$ & \textbf{0.8933}$\text{}^{\dagger \ddagger \flat}$ & \textbf{0.9400}$\text{}^{\dagger \ddagger \flat}$  \\
\textit{(g)} & \meta{} (RRF) & \textbf{0.7992}$\text{}^{\dagger \ddagger \flat \natural \sharp}$ & \textbf{0.8380}$\text{}^{\dagger \ddagger \flat \natural \sharp}$ & \textbf{0.7899}$\text{}^{\dagger \ddagger \flat \natural \sharp}$ & \textbf{0.8267}$\text{}^{\dagger \ddagger \flat \natural \sharp}$  & 0.8833$\text{}^{\dagger \ddagger \flat}$ & \textbf{0.9400}$\text{}^{\dagger \ddagger \flat}$ \\
\bottomrule
\end{tabular}
}

\caption{\label{tab:r5_result} Evaluation results of TREC-COVID R5. All Queries denotes all queries in R5. Old and New Queries denote queries that have been judged or not in previous rounds (R1-R4). (a) and (b) are the first-stage retrieval of other methods in this table. \textit{(c)} and \textit{(e)} are R5's top 2 automatic systems. \textit{(g)} is the reciprocal rank fusion (RRF) of \textit{(d)} and \textit{(f)}. $\dagger, \ddagger, \flat, \natural, \sharp, \S$ indicate statistically significant improvements over \textit{(a)}$\text{}^{\dagger}$, \textit{(b)}$\text{}^{\ddagger}$, \textit{(c)}$\text{}^{\flat}$, \textit{(d)}$\text{}^{\natural}$, \textit{(e)}$\text{}^{\sharp}$ and \textit{(f)}$\text{}^{\S}$.}
\end{table*}

% Evaluation results of TREC-COVID R5. All Queries denotes all queries in R5. Old and New Queries denote queries that have been judged or not in previous rounds (R1-R4). (a) and (b) are the first-stage retrieval of other methods in this table. \textit{(c)} and \textit{(e)} are R5's top 2 automatic systems. \textit{(g)} is the reciprocal rank fusion (RRF) of \textit{(d)} and \textit{(f)}. $\dagger, \ddagger, \flat, \natural, \sharp, \S$ indicate statistically significant improvements over \textit{(a)}$\text{}^{\dagger}$, \textit{(b)}$\text{}^{\ddagger}$, \textit{(c)}$\text{}^{\flat}$, \textit{(d)}$\text{}^{\natural}$, \textit{(e)}$\text{}^{\sharp}$ and \textit{(f)}$\text{}^{\S}$.}

%  \textit{(a)} and \textit{(b)} the base retrieval of (c) and (e).

% \textit{(a)} and \textit{(b)} are two baselines from Anserini BM25. \textit{(e)} and \textit{(c)} are the top 2 automatic teams in TREC-COVID R5 leaderboard. \textit{(d)} and \textit{(f)} use \textit{(a)} and \textit{(b)} as base runs respectively and are trained with MS MARCO + \contrast{} data. \textit{(g)} is the reciprocal rank fusion of \textit{(d)} and \textit{(f)}. Superscripts $\dagger, \ddagger, \flat, \natural, \sharp, \S$ indicate statistically significant improvements over \textit{(a)}$\text{}^{\dagger}$, \textit{(b)}$\text{}^{\ddagger}$, \textit{(c)}$\text{}^{\flat}$, \textit{(d)}$\text{}^{\natural}$, \textit{(e)}$\text{}^{\sharp}$ and \textit{(f)}$\text{}^{\S}$.
%% *********************************

%% *****************************
\begin{figure}[t]
\centering 
    \begin{subfigure}[t]{0.49\linewidth}
        \includegraphics[height=3.2cm]{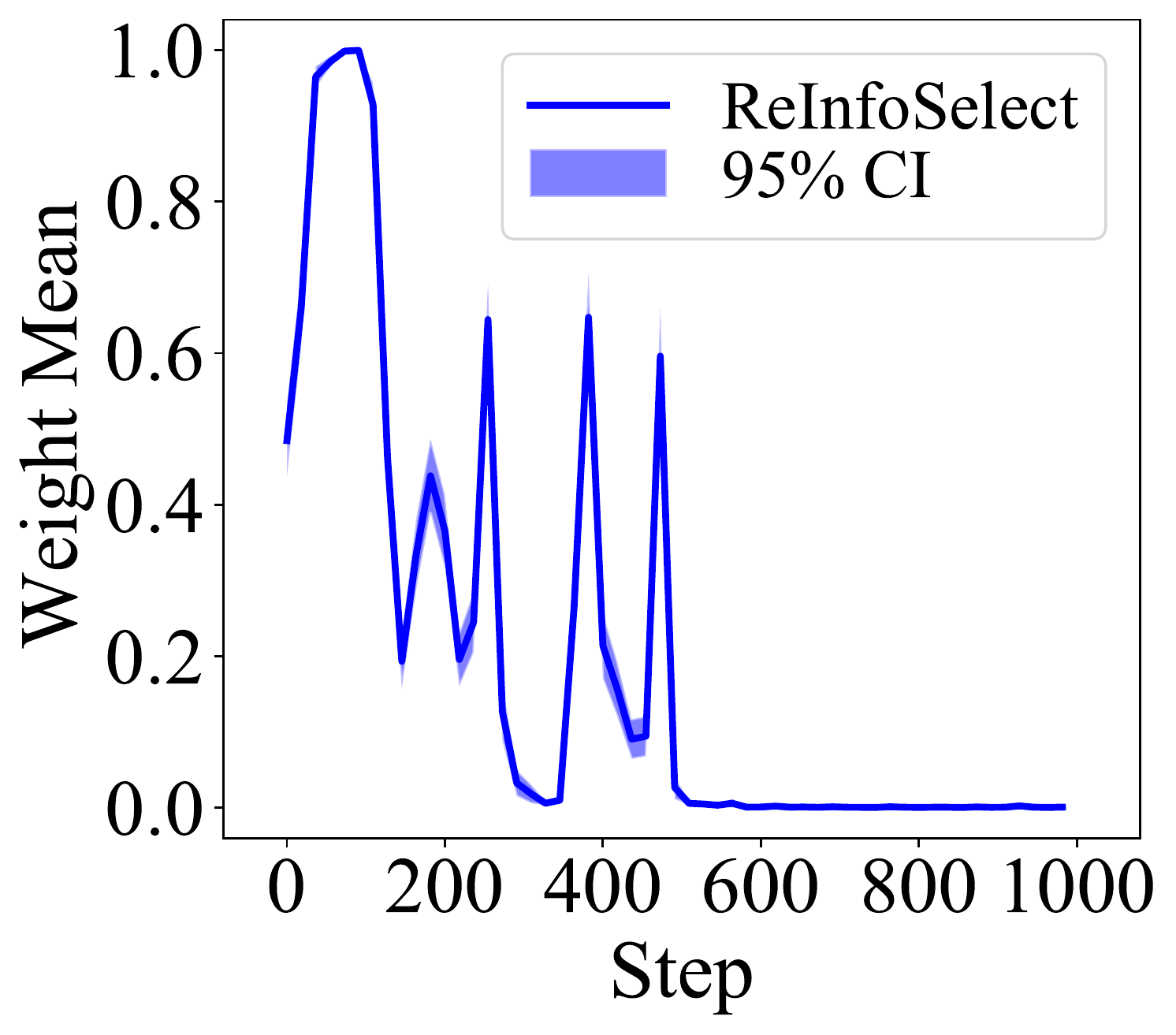}
        \caption{\rein{}.\label{fig:reinfo_weight}}
    \end{subfigure}
    \begin{subfigure}[t]{0.49\linewidth}
        \includegraphics[height=3.2cm]{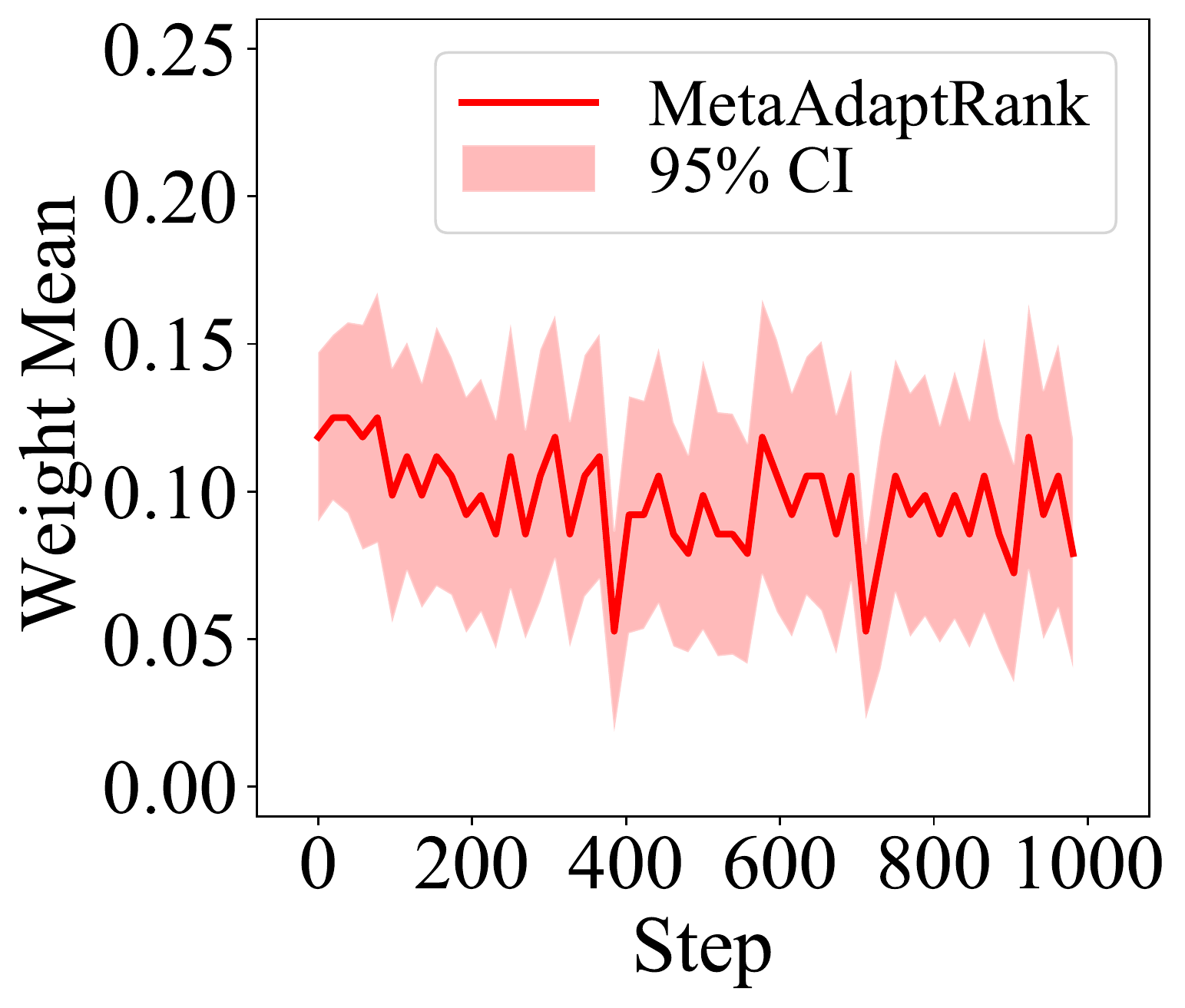}
        \centering
        \caption{\meta{}.\label{fig:meta_weight}}
    \end{subfigure}
\caption{\label{fig:weight_dist}The state of learned weights on \contrast{} data from ReInfoSelect and \meta{}. We use a ClueWeb09 few-shot fold as target data. Training steps are marked on X-Axes. The mean and 95\% Confidence Interval (CI) of data weights in the same batch are plotted. A 95\% CI is an interval that will contain the true mean of weights with 95\% probability. Its width is proportional to the standard deviation of data weights.
}
\end{figure}

\begin{figure}[ht]
\centering 
    \begin{subfigure}[t]{0.49\linewidth}
        \includegraphics[height=3.2cm]{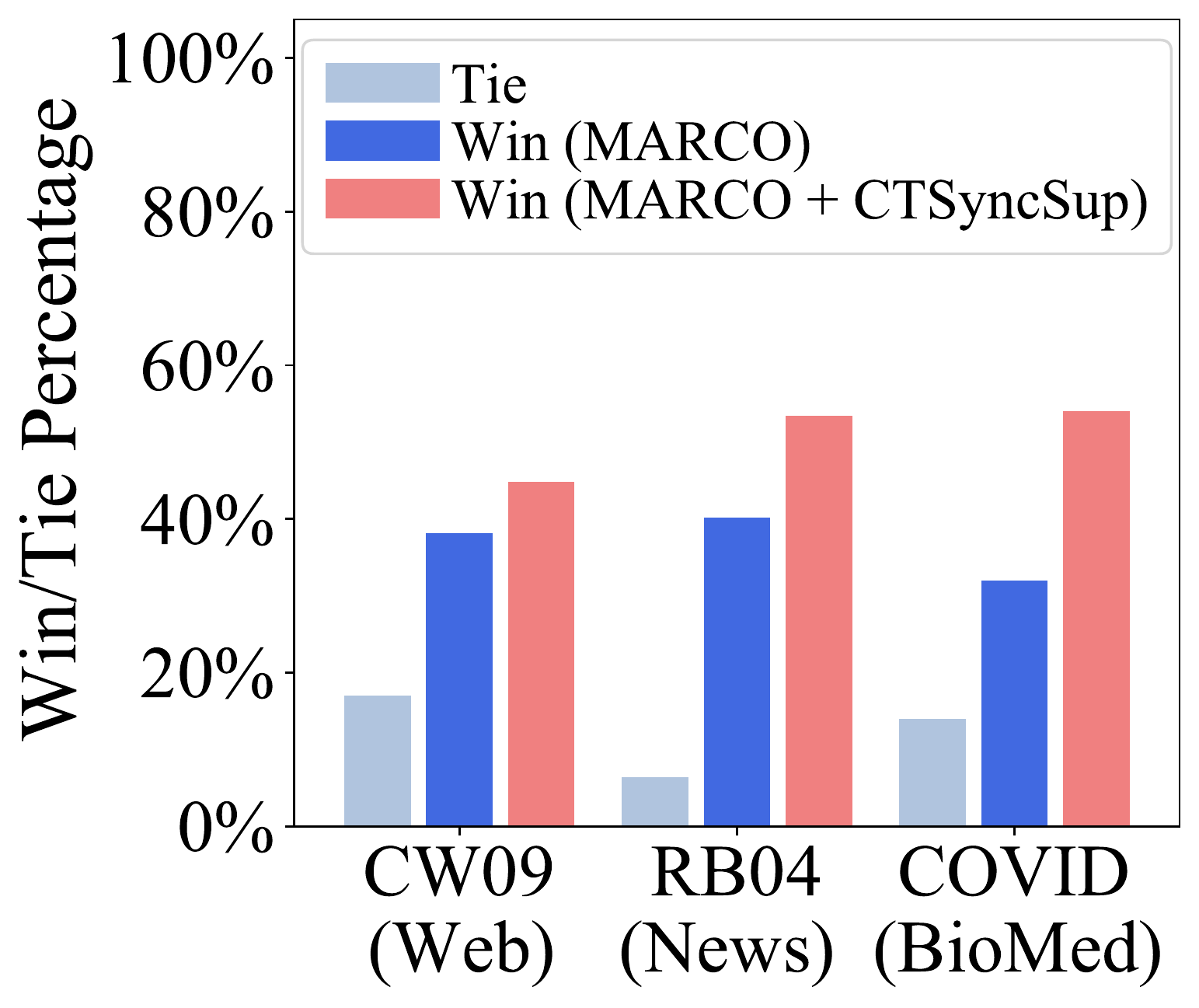}
        \caption{Win/Tie Percentage.\label{fig:wtl}}
    \end{subfigure}
    \begin{subfigure}[t]{0.49\linewidth}
        \includegraphics[height=3.2cm]{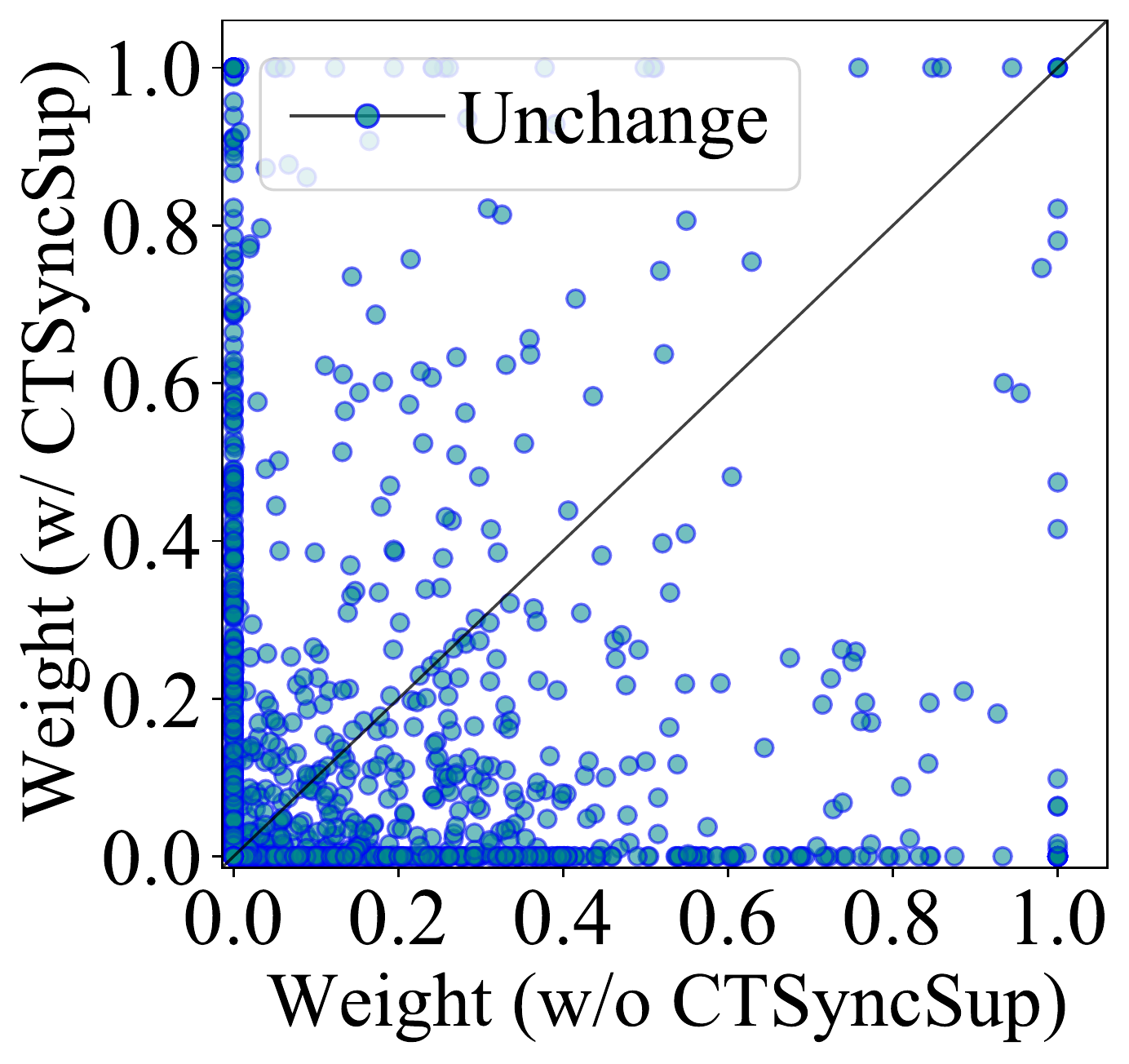}
        \caption{MARCO Weight Variation.\label{fig:marco_weight}}
    \end{subfigure}
\caption{\label{fig:analysis_fig} The analysis of \meta{} on the hybrid source of MS MARCO and \contrast{}. The ratio of Win/Tie queries between MS MARCO and the hybrid source is shown in (a). The statistics are based on NDCG@20 scores. CW09, RB04, COVID are short for datasets. (b) illustrates the variation in the meta-learned weights of 2k MS MARCO data points with (w/) and without (w/o) merging \contrast{}.
}
\end{figure}

\begin{table*}[t]
\centering
\small
\scalebox{0.9}{
\begin{tabularx}{\textwidth}{lX|X|X}
  \toprule
  \multicolumn{2}{c|}{\textbf{Synthetic Query}} & \textbf{Positive Document} & \textbf{Negative Document} \\ 
    \hline
    \multirow{2}{*}{($\uparrow$)} & 
    \texttt{\contrast{}:} how does \redbf{shopping with the planet make a big difference} in \redbf{msn} eco 
    \newline
    \texttt{\QG{}:} what is \bluebf{green energy} ecosystem
    & 
    … \bluebf{green} at \redbf{msn} shopping \redbf{shopping with the planet} in mind can \redbf{make a big difference} by \redbf{msn} shopping \redbf{msn} \bluebf{green} updated: \bluebf{energy} saving solutions conserving \bluebf{energy} reduces co2 emissions ...
    & 
    ... eco adventure tours \bluebf{energy} star pledge donate resources boater guide marinas harbormasters \bluebf{green} thumb ride share candle light dinner bright idea ...
    \\
    \hline
    \multirow{2}{*}{($\downarrow$)} & 
    \texttt{\contrast{}:} what is the \bluebf{history} of \redbf{bermuda}
    \newline
    \texttt{\QG{}:} where is \textit{jamestown beach}
    & 
    ... \redbf{bermuda} \textit{beach} resorts: website dedicated to advertising in \redbf{bermuda} large helpful travel forum \redbf{bermuda} links: activities hotels resorts \textit{beach} \ \redbf{bermuda} \bluebf{history} \redbf{bermuda} hotels ...
    & 
    ... your art \bluebf{history} reference guide art \bluebf{history} search \ \textit{jamestown}, colonial \bluebf{history} virginia (redirected from \textit{jamestown} settlement) \textit{jamestown} was a village on an island ...
    \\
    \bottomrule
\end{tabularx}
}
\caption{Cases of meta-reweighted contrastive synthetic data targeting ClueWeb09. The weights are marked in the parenthesis $\uparrow$ (more important) and $\downarrow$ (down-weight). The \redbf{red} texts are specific contents of positive documents and the \bluebf{blue} texts are shared by both positive and negative documents. The document snippets are manually selected.\label{tab:main_case}
}
\end{table*}

\subsection{Effectiveness of Meta Reweighting}
\label{sec:metaltr}

To analyze the effectiveness of meta reweighting, we employ \meta{} on different supervision sources and study its data weighting behaviors in the learning process. The reinforcement data selector \rein{} is used as a comparison, which utilizes the trial-and-error weighting mechanism.

The ranking accuracy of \meta{} and \rein{} trained with MS MARCO, Anchor, and \contrast{} is presented in Table~\ref{tab:adapt_ndcg}. For all supervision sources, \meta{} outperforms \rein{} on all benchmarks. The results show that the meta-reweighting mechanism can more effectively explore the potential of different supervision sources compared to the trial-and-error weighting mechanism. Moreover, the advantages of meta reweighting can be extended to the hybrid supervision source of MS MARCO and \contrast{}.

To further understand the behaviors of meta reweighting, we compare the state of weights assigned to synthetic supervision by \meta{} and \rein{} in the learning process, using \contrast{} as synthetic data and ClueWeb09 as target data. The results are shown in Figure~\ref{fig:weight_dist}. Even though each synthetic batch is likely to include both useful and noisy data points, \rein{} always assigns very high weights at the beginning and discards almost all synthetic data points later. Besides, its tight confidence interval reveals that data points in the same batch received almost identical weights. These observations indicate that \rein{} does not effectively distinguish useful synthetic data points from the noisy ones during the learning process. By contrast, {\meta{}} assigns higher weights initially and steadily reduces the weights as training goes on. More importantly, its wide confidence interval reveals that the data weights in the same synthetic batch vary significantly, which are thus expected to be more diverse and fine-grained.

% To further understand the meta-reweighting behaviors, we compare the state of synthetic supervision data weights assigned by \meta{} and \rein{} during the learning process, using \contrast{} as synthetic data and ClueWeb09 as target data. 

% To further understand the behaviors of meta reweighting, we compare the state of synthetic data weights learned by \meta{} with \rein{} in the learning process, using \contrast{} as synthetic weak supervision data and ClueWeb09 as target data. The results are shown in Figure~\ref{fig:weight_dist}. Even though each synthetic batch is likely to include both useful and noisy data points, \rein{} always assigns very high weights at the beginning and discards almost all synthetic data points later. Besides, its tight confidence interval reveals that data points in the same batch received almost identical weights. These observations indicate that \rein{} does not effectively distinguish the useful synthetic data points from the noisy ones during the learning process. By contrast, {\meta{}} assigns higher weights initially and steadily reduces the weights as training goes on. More importantly, its wide confidence interval reveals that the data weights in the same batch vary significantly, which are thus expected to be more diverse and fine-grained. 

%% *****************************************************
%% *****************************************************
\subsection{Effectiveness of Hybrid Supervision}
\label{sec:hybrid}

We also analyze \meta{}'s advantages on the hybrid supervision source of MS MARCO and \contrast{}. The impact of the hybrid source on its ranking accuracy and meta-reweighting behavior is studied. Besides, we evaluate \meta{} trained with the hybrid source in Round 5 of the TREC-COVID shared task in which many strong baselines have been well-tuned for four rounds.

Figure~\ref{fig:wtl} shows the Win/Tie ranking accuracy of \meta{} trained with MS MARCO and the hybrid supervision source. Compared to the single MS MARCO, the hybrid source has more advantages across all benchmarks. Besides, the hybrid advantage seems to be more evident in non-web domain benchmarks, especially on TREC-COVID. 

% Figure~\ref{fig:wtl} shows the Win/Tie ranking accuracy of \meta{} trained with MS MARCO and the hybrid supervision source. Compared to the single MS MARCO, the hybrid source has greater advantages across all benchmarks. Besides, the hybrid advantage seems to be more evident in non-web domain benchmarks, especially on TREC-COVID. 

We further investigate the weighting behavior of \meta{} on MS MARCO and the hybrid source, using the same ClueWeb09 target data in previous analyses. Figure~\ref{fig:marco_weight} illustrates the changes in meta-learned weights of randomly sampled 2k MS MARCO data points before and after merging \contrast{} source. There are significant weight variations on most MS MARCO data points before and after merging \contrast{}. Additionally, merging \contrast{} reduces the weight of more MS MARCO data points, revealing that \contrast{} data are assigned higher weights. This also reveals that \meta{} can tailor diversified weights for the same data points in different sources and up-weights more useful training data flexibly.

% We further investigate the weighting behavior of \meta{} on MS MARCO and the hybrid source, using the same ClueWeb09 target data in previous analyses. Figure~\ref{fig:marco_weight} illustrates the changes in meta-learned weights of randomly sampled 2k MS MARCO data points before and after merging \contrast{} source. There are significant weight variations on most MS MARCO data points before and after merging \contrast{}. Also, merging \contrast{} reduces the weight of more MS MARCO data points, revealing that \contrast{} data points are assigned higher weights. This further shows that \meta{} can tailor diversified weights for the same data points in different sources and up-weights more useful supervision data flexibly.

Lastly, we report the TREC-COVID R5 ranking results of \meta{} trained with the hybrid source. 
The top 2 automatic search systems in the R5 leaderboard are compared, which outperforms other systems on the newly added queries in R5. The evaluation of these new queries is fair to our methods and those systems that underwent previous rounds (R1-R4). As shown in Table~\ref{tab:r5_result}, our single model outperforms the top 2 fusion-based systems on all evaluation of the new, old, and all queries, further showing the effectiveness of \meta{} with the hybrid supervision source. More details and ranking results are shown in Appendices~\ref{sec:appendix:r5_new_query}.

\subsection{Case Studies}
\label{sec:case_study}

Table~\ref{tab:main_case} exhibits some cases of contrastive synthetic data for ClueWeb09 and their meta-learned weights. More cases are shown in Appendices~\ref{sec:appendix:case_study}.

\contrast{} can extract more specific contents from the positive documents, e.g., ``shopping with the planet'' and ``make a big difference'' in the first case; \QG{} captures more general information, e.g., ``green energy''. Compared to \QG{}'s queries such as ``where is jamestown beach'' in the second case, the synthetic queries in \contrast{} are more informative and discriminative. Noticeably, the second case exhibits the synthetic noise, where the positive document is actually related to ``bermuda's tourism'' instead of the query ``history of bermuda''. \meta{} effectively filters this noisy instance by assigning a zero weight to it.

\section{Conclusion}
This paper presents \meta{}, a domain adaption method for few-shot Neu-IR with contrastive weak data synthesis and meta-reweighted data selection. Contrastive synthesis generates informative queries and useful synthetic supervision signals. Meta-learned weights form high-resolution channels between target labels and synthetic signals, providing robust and fine-grained data selection for synthetic weak supervision. Both of them collaborate to significantly improve the neural ranking accuracy in various few-shot search scenarios.

\section*{Acknowledgments}
This work is partly supported by the National Key Research and Development Program of China (No. 2020AAA0106501) and Beijing Academy of Artificial Intelligence (BAAI). We thank Zhuyun Dai and Jamie Callan for sharing the SDM results on ClueWeb09-B and Robust04 and thank Shi Yu for discussions in the query generation methodologies.

% This work is partly supported by the National Key Research and Development Program of China (No. 2020AAA0106501) and Beijing Academy of Artificial Intelligence (BAAI). We thank Zhuyun Dai and Jamie Callan for sharing the SDM results on ClueWeb09-B and Robust04 and thank Shi Yu for discussions in query generation methodologies.

% Jimmy Lin and the Anserini project for open sourcing the well-rounded BM25 on TREC-COVID.

%% ********************************************
% \newpage
\balance
\bibliographystyle{acl_natbib}
\bibliography{anthology,acl2021}
\newpage
\appendix
\section{Appendices}

%% *****************************************
%% *****************************************
\subsection{Batch Normalization of Meta-Weights}
\label{sec:appendix:meta}
This part elaborates the batch normalization process for meta-learned weights. Following prior research~\cite{ren2018learning}, we first set the initial weights $w$ to zeros and obtain the new weights $\tilde{w}$:
\begin{equation}
    \small
    \tilde{w}_{j} = - \eta \frac{\partial  }{\partial  (w_j)}  \sum_{i=1}^{m} \frac{1}{m} l_i (\widetilde{\theta}^{t+1}(w))\Big|_{w_j=0}. \label{ref:eq:meta-gradient}
\end{equation}
Then we clip $\tilde{w}$ to get non-negative weights $\hat{w}$ and further normalize them in the batch to obtain the final weights $w^{*}$:
\begin{small}
\begin{equation}
\begin{split}
\hat{w}_{j} &= \max (0, \tilde{w}_{j}), \\
w_{j}^{*} &= \frac{\hat{w}_{j}}{(\sum_{p=1}^n \hat{w}_{p}) + \delta(\sum_{p=1}^n \hat{w}_{p})}.
\end{split}
\end{equation}
\end{small}
% \begin{equation}
% \begin{small}
% \begin{split}
% \hat{w}_{j} &= \max (0, \tilde{w}_{j}), \\
% w_{j}^{*} &= \frac{\hat{w}_{j}}{(\sum_{p=1}^n \hat{w}_{p}) + \delta(\sum_{p=1}^n \hat{w}_{p})}.
% \end{split}
% \end{small}
% \end{equation}
Here $\delta(\sum_{p=1}^n \hat{w}_{p}) = 1$ when $\sum_{p=1}^n \hat{w}_{p} = 0$, to prevent division errors, otherwise it is $0$. With the batch-normalization process, the hyperparameter $\eta$ can be effectively eliminated. The normalization method is not constrained and other approaches can also be used~\cite{shu2019meta, hu2019learning}.

%% *********************************
% \begin{figure*}[t]
%   \centering
%   \includegraphics[scale=0.31]{Figures/r5_new_query.pdf}
% \caption{\label{fig:r5_new_query} The NDCG@20 score on TREC-COVID R5 new queries (46-50) of top 20 leaderboard runs and our best two settings (\textit{f} and \textit{g} in table \ref{tab:r5_result}). Their order in the X-axis indicates the ranking of NDCG@20 score on all queries.}
% \end{figure*}

\begin{table}[t]
\centering
\small
\scalebox{0.94}{
\begin{tabular}{l l l}
\toprule
\textbf{Methods/Run ID} & \textbf{NDCG@20} & \textbf{P@20} \\ \hline
 UPrrf102-r5                     & 0.7873          & 0.8000          \\
UPrrf93-r5                      & 0.7967          & 0.8200          \\
covidex.r5.1s.lr                & 0.8019          & 0.8300          \\
elhuyar\_prf\_nof99d            & 0.8209          & 0.8700          \\
covidex.r5.d2q.1s.lr            & 0.8287          & 0.8400          \\
elhuyar\_prf\_nof99p            & 0.8333          & 0.9000          \\
MetaAdaptRank (rerank fusion.1) & 0.8712 & 0.9300 \\
UPrrf102-wt-r5                  & 0.8804          & 0.9100          \\
MetaAdaptRank (RRF)             & 0.8833          & \textbf{0.9400}\\
covidex.r5.2s.lr                & 0.8837          & 0.9000          \\
UPrrf93-wt-r5                   & 0.8849          & 0.9100          \\
covidex.r5.d2q.2s.lr            & 0.8929          & 0.9100          \\
MetaAdaptRank (rerank fusion.2) & \textbf{0.8933} & \textbf{0.9400} \\
\bottomrule
\end{tabular}
 }
  \caption{\label{tab:r5_new_query}Ranking results of our methods and baselines on the new queries of TREC-COVID R5. The baselines are the top 10 feedback systems in the R5 leaderboard, marked with their submitted ID. The three variants of \meta{} are the same as those in Table~\ref{tab:r5_result}.}
\end{table}

%   The TREC-COVID R5 new queries results of \meta{} and top 10 feedback systems in R5 leaderboard. \meta{} (rerank fusion.1), \meta{} (rerank fusion.2) and \meta{} (RRF) keep the same settings with the models in Table~\ref{tab:r5_result}.
%   \caption{\label{tab:r5_new_query} TREC-COVID R5 new queries (46-50) results of top 10 leaderboard runs and three of our settings. \meta{} (rerank fusion.1), \meta{} (rerank fusion.2) and \meta{} (RRF) are our settings (the same with those in Table \ref{tab:r5_result}), and the others are feedback runs from TREC-COVID R5 leaderboard.} 
%% *********************************
%% *********************************
\begin{table*}[ht]
\centering
\small
\scalebox{0.93}{
\begin{tabularx}{\textwidth}{lX|X|X}
  \toprule
  \multicolumn{2}{l|}{\textbf{Synthetic Query}} & \textbf{Positive Document} & \textbf{Negative Document} \\ 
    \hline
    %% ********************************************
    %% Robust04 (1+)
    \multirow{2}{*}{1 ($\uparrow$)} &
    \texttt{\contrast{}:} us military \redbf{radars} in \redbf{colombia} 
    \newline
    \texttt{\QG{}:} what is the \bluebf{pentagon}
    & 
    ... one month ago, the \bluebf{pentagon} issued an order to suspend operations of the two \redbf{radars} that detect aircraft. these \redbf{radars} operate in \redbf{colombia} as a result of that agreement ...
    & 
    ... provide for more funding and retain more forces than the \$1.5-trillion five-year budget cheney presented to congress in january, \bluebf{pentagon} officials say ...
    \\
    \hline
    %% ********************************************
    %% Robust04 (2+)
    \multirow{2}{*}{2 ($\uparrow$)} &
    \texttt{\contrast{}:} what percent of the economy was \redbf{increased} in \redbf{1993}
    \newline
    \texttt{\QG{}:} what is the economic issue in \bluebf{peru}
    & 
    ... this letter explains the \bluebf{peruvian} government's economic policy. \  the development of the economy in \redbf{1993} was in general much better. \  it is estimated that the real gdp has \redbf{increased} by 7 percent  ...
    & 
    ... only three economies - guyana, argentina and \bluebf{peru} - grew by more than than 5 per cent this year, with \bluebf{peru} expanding by 11 percent ...
    \\
    \hline
    %% ********************************************
    %% Robust04 (3-)
    \multirow{2}{*}{3 ($\downarrow$)} &
    \texttt{\contrast{}:} what \redbf{language} is \bluebf{osvaldo rodriguez}
    \newline
    \texttt{\QG{}:} what is economic impact of cuba
    & 
    ... program with host \bluebf{osvaldo rodriguez} \ the dialogue that was proposed by opponents to the revolutionary project. well, the word dialogue, is one that connotes cordiality. it is a positive word. but in the current political \redbf{language}, the counterrevolution's political \redbf{language} ...
    & 
    ... program with host juan carlos roque garcia, and \bluebf{osvaldo rodriguez}. this program could not ignore cuba's presentation of a document by the u.s. interests section in havana to the un human rights commission in geneva ...
    \\
    \hline
    \hline
    %% ********************************************
    %% TREC-COVID (1)
    \multirow{2}{*}{4 ($\uparrow$)} &
    \texttt{\contrast{}:} which receptors are expressed in the human \redbf{lung}
    \newline
    \texttt{\QG{}:} what is \bluebf{sars cov} receptor
    & 
     ... results both \bluebf{sars-cov} receptors of ace2 and cd209l were expressed in the 8 organ/tissue-derived endothelial cells. the expression of ace2 receptor was the highest in the human \redbf{lung} microvascular endothelial cells, and lowest ...
    & 
    ... 2019 novel coronavirus (2019-n\bluebf{cov}) the outbreaks of 2002/2003 \bluebf{sars}, 2012/2015 mers and 2019/2020 wuhan respiratory syndrome clearly indicate that genome evolution of an animal coronavirus (\bluebf{cov}) may enable ...
    \\
    \hline
    \multirow{2}{*}{5 ($\uparrow$) } &
    \texttt{\contrast{}:} how does \redbf{quarantine prevent} covid outbreak
    \newline
    \texttt{\QG{}:} covid outbreak \bluebf{symptoms}
    & 
     ... the importance of the timing of \redbf{quarantine} measures before \bluebf{symptom} onset to \redbf{prevent} covid-19 outbreaks \ how \redbf{quarantine}-based measures can \redbf{prevent} or suppress an outbreak ...
    & 
    ... furthermore, the effect of infectiousness prior to \bluebf{symptom} onset combined with a significant proportion \ we evaluate two procedures: monitoring individuals for \bluebf{symptoms} onset ...
    \\
    \hline
    \multirow{2}{*}{6 ($\downarrow$)} &
    \texttt{\contrast{}:} \bluebf{covid-19 pandemic} effects on society
    \newline
    \texttt{\QG{}:} what is the antiasia \textit{sentiment} in the united states
    & 
    ... examination of community \textit{sentiment} dynamics due to \bluebf{covid-19 pandemic}: the outbreak of covid-19 has caused unprecedented impacts to people's daily life around the world. virus may cause different mental health issues to people such as depression, anxiety, sadness ...
    & 
    ... mood of india during covid-19 - an interactive web portal based on emotion analysis of twitter data the \bluebf{covid-19 pandemic} has affected many countries across the world, and disrupted the day to day activities of many people ...
    \\

    \bottomrule
\end{tabularx}
}
\caption{The contrastive synthetic data reweighted by \meta{}, where the top 3 cases are from Robust04 (News) and the last 3 cases come from TREC-COVID (BioMed). Their meta-weights are marked in the parenthesis $\uparrow$ (more important) and $\downarrow$ (down-weight). The \redbf{red} texts are the specific contents of the positive documents, and the \bluebf{blue} texts are mentioned in both positive and negative documents. The document snippets are manually selected.\label{tab:appendix_case}
}
\end{table*}

% The synthetic supervision signals reweighted by \meta{}, where the top 3 cases are from Robust04 (News) and the last 3 cases come from TREC-COVID (BioMed). The directions of their meta-weights are marked in the parenthesis $\uparrow$ (more important) and $\downarrow$ (down-weight). The \bluebf{blue} texts are shared by the two documents, and the \redbf{red} texts are the specific contents of the positive document. The document snippets are manually selected
%% *********************************

%% *****************************************
%% *****************************************

\subsection{Supplementary Results of TREC-COVID R5}
\label{sec:appendix:r5_new_query}

This part supplements our evaluation results in the TREC-COVID R5 shared task. We will first recap the shared task and then present more evaluation results and our implementation details.

\textit{TREC-COVID R5.} The TREC-COVID Challenge is an ad-hoc ranking task for COVID-19 literature, consisting of five rounds. TREC-COVID R5 is the last round of this challenge, where the document set is the July 16, 2020 version of CORD-19, and the query set contains 50 testing queries. The first 45 queries have been used in previous rounds (R1-R4), and the last five queries are newly added in R5. As in previous rounds, TREC-COVID R5 adopts residual collection evaluation~\cite{Salton1997ImprovingRP}. In residual collection evaluation, the relevance labels from previous rounds can be used, but any document that has been annotated for a query will be removed before the evaluation. We focus more on the evaluation of R5's new queries because these queries have no prior relevance labels, which is fairer to our models and those search systems that underwent previous rounds.

\textit{Evaluation Results.} Table~\ref{tab:r5_new_query} shows the evaluation results on the new queries of TREC-COVID R5, including three variants of our \meta{} and the top 10 feedback systems in the R5 leaderboard. Compared with the top 10 feedback systems (many are fusion-based systems), our single model \meta{} (rerank fusion.2) outperforms all baselines, demonstrating the generalization ability of our method on new queries.

Additionally, what catches our attention is that the best and worst of the top 10 feedback systems only have a 5.1\% difference in NDCG@20 scores on all queries, while their NDCG@20 scores on the new queries differ by 13.4\%. This discrepancy indicates that the residual collection evaluation may have biases between the seen and unseen queries.

\textit{Implementation Details.} We next describe the implementation details of the three variants of our \meta{} in TREC-COVID R5. Consistent with the implementation methods described in Section~\ref{sec:method}, we rerank the top 100 documents from the first-stage retrieval. We first borrow two retrieval results with different settings provided by Anserini BM25 (Row 7 and 8 of Table Round 5\footnote{\url{https://github.com/castorini/anserini/blob/master/docs/experiments-covid.md}}). Then PudMedBERT (Base) is used to rerank these two retrieval results to obtain \meta{} (rerank fusion.1) and \meta{} (rerank fusion.2), respectively. \meta{} (RRF) is the reciprocal rank fusion of these two models. We utilize the open-source library trec-tools~\cite{palotti2019} to implement RRF and set the fusion weight $k$ to 1.

To train \meta{}, we first synthesize \contrast{} data based on R5's document set and leverage the hybrid source of \contrast{} and MS MARCO as the additional supervision signals. The training process contains two stages. We first train \meta{} with the hybrid source and regard the labeled data from previous rounds (R1-R4) as target data in meta-reweighting. Then we continuously train \meta{} using the labeled data from the previous rounds. In the training processes, we utilize Adam optimizer with a learning rate of 2e-5. Both the batch size and the accumulation step are set to 8. In addition, to ensure a fair comparison with the submitted search systems, we post-process our results according to official guidelines.

%% *****************************************
%% *****************************************

\subsection{Supplementary Case Studies}
\label{sec:appendix:case_study}

Table~\ref{tab:appendix_case} shows more cases for the other two datasets, Robust04 (News) and TREC-COVID (BioMed), to verify the effectiveness of \meta{} in different domains. The first three cases are from Robust04, and the rest cases are from TREC-COVID.

For the first synthetic cases, our \contrast{} can extract characteristic keywords, e.g., ``radars'' and ``colombia'', from the positive documents to generate more informative queries, while \QG{} tends to capture general keywords to create broad queries, which may lack the ability to distinguish between different documents. Besides, \contrast{} can extract some necessary themes from the specific documents, such as the particular time ``1993'' and the adjective ``increased'', as shown in the second case.

Moreover, cases 4 and 5 show the effectiveness of our contrastive synthesis for biomedical domains. \contrast{} can capture ``lung'' and ``quarantine prevent'' instead of general keywords, such as ``sars cov'' and ``symptoms'' often mentioned in COVID-related documents. These observations show that \contrast{} can extract more specific information to generate more informative and discriminative queries for different target domains.

We further explore those synthetic instances that are assigned zero weights by \meta{}, such as the third and sixth cases. In the third case, although \contrast{} captures the two keywords ``language'' and ``osvaldo rodrigrez'' from the positive document, its synthetic query is actually less relevant to the main topic of the positive document. For the sixth case, \contrast{} fails to exclude the phrase ``covid-19 pandemic'' related to both the positive and negative documents, which causes the synthetic query unable to distinguish between them. Fortunately, \meta{} can effectively identify the synthetic instances whose relevance matching patterns between synthetic queries and positive documents are unclear or non-unique and then precludes such misleading synthetic supervision data by assigning them zero weights.

\end{document}